# Magnetotransport studies of optimally doped Sr(Fe$_{1-x}$Co$_x$)$_2$As$_2$

Rohit Kumar[1], Luminita Harnagea[1], Archana Lakhani[2] and Sunil Nair[1,3]

[1]Department of Physics, Indian Institute of Science Education and Research, Dr. Homi Bhabha Road, Pune, Maharashtra-411008, India

[2]UGC-DAE Consortium for Scientific Research, University Campus, Khandwa Road, Indore-452017, Madhya Pradesh, India

[3]Centre for Energy Science, Indian Institute of Science Education and Research, Dr Homi Bhabha Road, Pune, Maharashtra-411008, India

## Abstract

We report magnetotransport measurements and its scaling analysis for the optimally electron doped Sr(Fe$_{0.88}$Co$_{0.12}$)$_2$As$_2$ system. We observe that both the Kohler's and modified Kohler's scalings are violated. Interestingly, the Hall angle displays quadratic temperature dependence ($cot\theta_H \propto T^2$) similar to many cuprates and heavy fermion systems. The fact that this T$^2$ dependence is seen in spite of the violation of modified Kohler's scaling suggests that the Hall angle and magnetoresistance are not governed by the same scattering mechanism. We also observe a linear magnetoresistance in this system, which does not harbour a spin density wave (SDW) ground state. Implications of our observations are discussed in the context of spin fluctuations in strongly correlated electron systems.

## I. Introduction

Copper oxide superconductors were the only high-$T_C$ materials until 2008, when superconductivity at ≈ 26 K was discovered in a fluorine doped oxypnictide La(O$_{1-x}$F$_x$)FeAs [1]. Since then, many families have been discovered, most of them having FeX (X = As or P etc.) layers as a common ingredient, which is crucial to the physics of these systems. Most studied among these are the so called 122 families, mostly due to the availability of relatively large and pure single crystals. Parent compounds of these families are multiband semimetals [2, 3, 4]. One of the most interesting property is the transport behaviour of these materials: temperature dependent Hall coefficient [5, 6], quasi linear temperature dependent resistivity [7] and linear magnetoresistance [8-16] in the Spin Density Wave (SDW) state to name a few. Despite the fact that single crystals of Sr(Fe$_{1-x}$Co$_x$)$_2$As$_2$ with lateral dimensions upto several millimetres can be grown, magnetotransport measurements are still lacking in this series, especially in its electron doped variants.

Sr(Fe$_{1-x}$Co$_x$)$_2$As$_2$ has few unique features among the 122 family of iron based superconductors (FeSC). For example, magneto-structural transitions do not split upon doping [17, 18], in sharp contrast to some other families of the FeSCs [19, 3, 20]. Interestingly, despite having small values of Residual Resistivity Ratio, which are very similar ( ~ 3) to other members of the 122 families of FeSCs, quantum oscillations in the resistivity have been observed at magnetic fields as low as 12 Tesla [21, 22] and mean free paths as large as 2700 Å were reported [22].

It is well known that structural details are crucial to the physics of iron based superconductors [3, 23, 4]. For example, it was pointed out that the maximum $T_C$ in many families of the iron pnictides occurs when the FeAs$_4$ tetrahedron is least distorted i.e., the As-Fe-As bond angle in the FeAs$_4$ tetrahedron is close to its ideal value of 109.47˚ [3, 23]. It was proposed that for the spin fluctuation mediated superconductivity, the height of pnictogen atoms above the Fe sublattice ($h_{pn}$) is also an important determining factor for not only $T_C$ but also for the gap symmetry [24]. It was suggested that the superconductivity and magnetism are controlled by an intricate balance between the $h_{pn}$ and lattice constants [24]. For example, SrFe$_2$As$_2$ and EuFe$_2$As$_2$ have very similar lattice parameters and ionic radii, and consequently their magnetostructural transition temperatures and some other properties are very similar [25, 26]. This is in contrast to the BaFe$_2$As$_2$ system, where larger lattice constants and higher ($h_{pn}$) leads to lower magnetostructural transition temperatures and higher superconducting transition temperatures.

We report magnetotransport measurements and its scaling analysis in an optimally doped (x = 0.12) composition of the Sr(Fe$_{1-x}$Co$_x$)$_2$As$_2$ series. Scaling analysis is a powerful tool to find commonalities and differences in different classes of materials. In the past, many cuprates and heavy fermion systems were shown to have similar magnetotransport scaling behavior, indicating that underlying mechanism behind their unconventional properties might be similar [27-29]. When, the transport is dominated by only one type of charge carrier, relaxation rate is same across the entire Fermi surface, and the carrier concentration do not change with temperature, then the magnetoresistance ($\Delta\rho/\rho_0$) can be scaled as $(H/\rho_{xx}(0))^2$, where H is the applied magnetic field, $\rho_{xx}(0)$ is the zero field resistivity and $\Delta\rho$ is the change in resistivity after the application of magnetic field. This is called Kohler's rule [30, 31, 32]. Kohler's rule is often violated in the strange metal phase of high-$T_C$ cuprates, heavy fermions and iron based superconductors [7, 28, 33-37].

In many cuprates and heavy fermions, magnetoresistance was shown to scale not by $(H/\rho_{xx}(0))^2$, but by $tan\theta_H$ where $tan\theta_H = \rho_{xy}/\rho_{xx}$ [34]. This modified Kohler's scaling was shown to be valid in a number of heavy fermion systems [27-29, 38] and some FeSCs as well [7, 36, 35], indicating that the Magnetotransport in these systems is governed by the same mechanism and antiferromagnetic fluctuations are thought to be at the origin of these unconventional properties [29, 27]. We observe that both the Kohler's and modified Kohler's scaling do not work for the $Sr(Fe_{0.88}Co_{0.12})_2As_2$ system. Another interesting observation is of a quadratic temperature dependence of the cotangent of the Hall angle ($cot\theta_H \propto T^2$) reminiscent of many Cuprates and heavy fermion superconductors [34, 27, 28, 39, 40, 41]. It is known that, parent compound $SrFe_2As_2$, displays linear magnetoresistance in the SDW state [10, 42, 43] (along with small quadratic contribution at low fields), like many other members of FeSCs [8 - 16]. In the present study, we also report linear magnetoresistance in the strange metal phase of the optimally doped $Sr(Fe_{0.88}Co_{0.12})_2As_2$ system.

## II.   Materials and Methods

Single crystals of $Sr(Fe_{1-x}Co_x)_2As_2$ (x = 0.12) were grown using FeAs as flux. Here, x = 0.12 is actual composition of the crystals determined by Energy Dispersive X-Ray Spectroscopy (EDX). Details of single crystal growth and characterization can be found elsewhere [44]. Magnetotransport measurements were carried out in a 9 Tesla Quantum Design PPMS. Electrical contacts on the sample surface were made using a gold wire of 25 micron diameter and silver epoxy. To remove any exfoliating layers, samples were slightly polished before the measurements. Magnetic field was applied along the crystallographic c axis and electrical current in the ab plane. Polarity of the magnetic field was reversed at each measurement and the Hall resistivity was extracted as the asymmetric component of the signal i.e. $\rho_{xy} = [\rho(+H) - \rho(-H)]/2$. Since magnetoresistance in these systems is very small at high temperatures, especially in the paramagnetic phase, averaging routines were employed to enable the scaling analysis.

## III. Results and Discussion

Main panel of Figure 1 shows the in-plane resistivity from 2-300 K. There are no discernible anomalies associated with the magnetic/structural transitions, which are present in the under doped systems, implying that the magnetic and structural transitions are completely suppressed. Inset shows the resistivity near the superconducting region. $T_C$ is assigned to the midpoint of the transition ($\approx$14.5 K), which is approximately the average of onset and offset temperatures which are shown by the arrows. Similar values of optimal $T_C$ were reported previously [18, 45, 44]. Interestingly, maximum $T_C$ in the as grown crystals corresponding to the optimally doped composition (x ~ 0.07) in the Ba(Fe$_{1-x}$Co$_x$)$_2$As$_2$ series is ~ 25 K [46, 47, 19]. The maximum $T_C$ corresponding to the optimal Co doping in as grown Sr(Fe$_{1-x}$Co$_x$)$_2$As$_2$ series is smaller by approximately 10 K.

This difference between the maximum $T_C$ of Ba(Fe$_{1-x}$Co$_x$)$_2$As$_2$ and Sr(Fe$_{1-x}$Co$_x$)$_2$As$_2$ series can be understood from the fact that maximum $T_C$ is expected to occur in systems with low effective dimensionality. This is because the strength of spin fluctuations is stronger in low dimensional systems and this can lead to higher $T_C$. For instance, it is known that the 1111 systems are most 2D in nature compared to any other member of the FeSCs [48], as a consequence of which, highest $T_C$ are observed in the 1111 systems [3]. Sr(Fe$_{1-x}$Co$_x$)$_2$As$_2$ system is less two dimensional in nature as compared to Ba(Fe$_{1-x}$Co$_x$)$_2$As$_2$ system. For example, the ratio of the in-plane and out of plane plasma frequencies ($\omega_a^p/\omega_c^p$), which can serve as a quantitative measure of the effective dimensionality was shown [48] to be higher in BaFe$_2$As$_2$: $\omega_a^p/\omega_c^p$ = 3.29 for BaFe$_2$As$_2$ and $\omega_a^p/\omega_c^p$ = 2.83 for SrFe$_2$As$_2$. This can explain why the optimum $T_C$ is lower in the Sr(Fe$_{1-x}$Co$_x$)$_2$As$_2$ series as compared to the Ba(Fe$_{1-x}$Co$_x$)$_2$As$_2$ series.

Interestingly, the critical concentration (x$_c$) where optimal $T_C$ is observed, ($x_c \approx 0.12$ in the present case) is much larger in the Sr(Fe$_{1-x}$Co$_x$)$_2$As$_2$ series as compared to the Ba(Fe$_{1-x}$Co$_x$)$_2$As$_2$ series ($x_c \sim 0.07$). This might be due to the fact that the magnetostructural transition occurs at ~ 200 K in Sr(Fe$_{1-x}$Co$_x$)$_2$As$_2$ series which is quite high as compared to the Ba(Fe$_{1-x}$Co$_x$)$_2$As$_2$series. It appears that in order to obtain optimal $T_C$ in Sr(Fe$_{1-x}$Co$_x$)$_2$As$_2$ series, more carriers need to be doped. Similar results were also reported for the BaFe$_2$(As$_{1-x}$P$_x$)$_2$ [49] and (Ba$_{1-x}$K$_x$)Fe$_2$As$_2$ [50] systems, where the optimal $T_C$ is obtained only after the complete suppression of the magnetostructural transitions.

Figure 2(a) shows the Hall resistivity ($\rho_{xy}$) as a function of the applied magnetic field. Hall resistivity is seen to be linear in the magnetic field which is similar to the optimally doped concentration of the Ba(Fe$_{1-x}$Co$_x$)$_2$As$_2$ series [5]. This allows for an unambiguous determination of the Hall coefficient, which is shown in Figure 2(b). The Hall coefficient is strongly temperature dependent which is reminiscent of many other iron pnictides [51, 52, 5] and strongly correlated electron systems [53, 27].

Figure 3(a) and (b) shows Kohler's and the modified Kohler's plots respectively. Magnetoresistance curves do not collapse on top of each other implying that both scalings are violated in the entire temperature range. As stated previously, Kohler's rule reads: $\Delta\rho/\rho_0 \propto (H/\rho_{xx}(0))^2$. In the Drude picture, it can be written as: $\Delta\rho/\rho_0 \propto mH/ne^2\tau$, where m is the electronic mass, $\tau$ is the relaxation time and $H$ is the applied magnetic field. Kohler's rule is found to be valid for a number of simple metals [31, 32] and even in over doped regime of strongly correlated electron systems like FeSCs, cuprates and heavy fermions where Fermi liquid like behaviors are typically recovered [33, 36, 27]. Violation of Kohler's rule in strongly correlated electron systems, especially in their strange metals phase is a common occurrence [7, 34, 29]. This is because the premise on which it is based, i.e., of a single species of charge carrier dominating the transport, relaxation time being invariant over the Fermi surface and carrier concentration not changing with temperature, are often not met in these systems [54].

Unfortunately, to pin down the exact reason for the violation of Kohler's rule is also not straight forward, especially in complex systems such as the FeSCs because any one or more of the above mentioned factors may be responsible for the violation. For example, it is well known that FeSCs are multiband systems [2, 4], a situation not favourable for Kohler's scaling. Also, the Hall coefficient is known to be strongly temperature dependent [5, 6], which would mean that carrier concentration is not constant with temperature, which is again an unfavourable condition for the Kohler's scaling. As stated previously, we also see strongly temperature dependent Hall coefficient, see Figure 2(b). In fact in Co doped BaFe$_2$As$_2$ systems, carrier concentration was shown to be strongly temperature dependent in an ARPES study [55]

Another possible reason for the violation of the Kohler's rule could be the variation of relaxation time across the Fermi surface. Such an anisotropic reconstruction of the Fermi surface is reported in FeSCs [56-58] and similar formalisms have been used to explain many non-Fermi liquid like behaviors of the high-$T_C$ copper oxide superconductors [53, 59-61]. To account for the anomalous transport properties of cuprates, existence of two relaxation times was proposed. It was suggested that the resistivity is governed by a transport relaxation time ($\tau_{tr}$) and $cot\theta_H$ is governed by the so called Hall relaxation time $\tau_H$ [59, 34]. In many cuprates, $\rho_{xx}$ is linear in temperature, at least in some region of the temperature-composition phase space whereas $\rho_{xy}$ varies as 1/T. As mentioned earlier, the modified Kohler's rule states that MR $\propto tan^2\theta_H$ or $1/cot^2\theta_H$ Since, both the magnetoresistance and $cot\theta_H$ are determined by the Hall scattering time $\tau_H$ [59, 34], T$^2$ dependence of $cot\theta_H$ and the validity of modified Kohler's scaling are often taken as the validation of this theory. This implies that if the modified Kohler's scaling is invalid, Hall angle should not be quadratic in temperature in this picture. This was observed in a number of cuprates and heavy fermion systems [28, 34, 38, 29]. Some isovalently doped FeSCs were also shown to obey the modified Kohler's scaling. [36, 7]. The plot of $cot\theta_H$ as a function of $T^2$ for our system is shown in Figure 4 . Evidently, a good fit is obtained in most of the temperature range; however the fit begins to deviate from $T^2$ behavior at low temperatures around 40 K. We suspect that this deviation is due to the proximity to superconducting transition. Note that even in some high-$T_C$ cuprates, modified Kohler's scaling was shown to be invalid even when $cot\theta_H$ had quadratic temperature dependence [39, 40, 62, 63]. Consequently, it was suggested that modified Kohler's rule is not universally applicable to all high-$T_C$ cuprates either [63, 39, 40]. Our results are the first in iron-based superconductors to suggest the same.

Magnetotransport behavior of the electron doped 122 families appears to be different from that of the isovalently doped systems. For example, it is known that the parent compound $BaFe_2As_2$ and the optimally doped composition (x = 0.074) of the $Ba(Fe_{1-x}Co_x)_2As_2$ series do not obey modified Kohler's scaling [64, 12]. On the other hand isovalently doped optimal composition of the $BaFe_2(As_{1-x}P_x)_2$ [7] and $Ba(Fe_{1-x}Ru_x)_2As_2$ [36] series were shown to obey the modified Kohler's scaling. It should be noted that under doped composition of $Ba(Fe_{1-x}Co_x)_2As_2$ series is an interesting exception here [64, 12]. Modified Kohler's rule was found to be obeyed in the SDW state of two different electron under doped compositions [64, 12]. These observations imply that the scenario of separation of scattering times may not be applicable to the electron doped 122 families of $Ba(Fe_{1-x}Co_x)_2As_2$ and $Sr(Fe_{1-x}Co_x)_2As_2$ FeSCs but is applicable to 122 families of isovalently doped iron pnictides.

It has been argued that many unconventional transport properties of the high-$T_C$ cuprates like strongly temperature dependent Hall coefficient, modified Kohler's rule etc. can be derived within the framework of nearly antiferromagnetic Fermi liquid if the current vertex corrections are taken into account [53, 65-67]. In this theory, the Hall coefficient and magnetoresistance are both normalized due to the temperature dependence of the antiferromagnetic correlation length ($\xi_{AF}$) i.e., $R_H \propto \xi_{AF}^2$ and $\Delta\rho/\rho_0 \propto \xi_{AF}^4 H^2/\rho_0^2$ Evidently, from these expressions, Kohler's rule is violated in the presence of strongly temperature dependent $\xi_{AF}$ whereas the modified Kohler's rule $\Delta\rho/\rho_0 \propto R_H^2/\rho_0^2$ remains valid. However, as we can see from Figure 3(b) modified Kohler's scaling is not valid in the system under study. Very similar scaling behavior of Ba(Fe$_{1-x}$Co$_x$)$_2$As$_2$ and Sr(Fe$_{1-x}$Co$_x$)$_2$As$_2$ however require a coherent description of normal state transport properties of electron doped FeSCs.

We now turn our attention to the phenomena of linear magnetoresistance (LMR). As is evident from Figure 5, magnetoresistance is linear in magnetic field similar to what was seen in the paramagnetic phase of electron doped Ba(Fe$_{1-x}$Co$_x$)$_2$As$_2$ systems [64]. As mentioned previously, linear magnetoresistance has been observed in SDW state of many families of FeSCs [8-16]. It is often assumed to originate due to the presence of Dirac cone states which arises due to the reconstruction of the Fermi surface that occurs at the onset of SDW instability [8, 15, 11]. Dirac cone states are indeed observed in the photoemission [68] and quantum oscillation experiments [22, 69] and are now an established fact in FeSCs.

This linear magnetoresistance is often explained using the quantum linear magnetoresistance (QLM) model of Abrikosov [70-72]. In this model, linear magnetoresistance was predicted in the quantum limit where all carriers occupy the lowest Landau band. Thus, $\rho_{xx} \propto NH/n^2$ provided $n \ll (eH/c\hbar)^{3/2}$ and $T \ll eH\hbar/m^*$, here $N$ and $n$ are the density of scattering centers and charge carriers respectively and $H$ is the applied magnetic field. Clearly, low carrier concentration, low temperature and high magnetic field are the favourable conditions to obtain this quantum limit. It is believed that the quantum limit can be reached even at relatively high temperatures and typical laboratory magnetic fields in small Dirac pockets which are formed in the SDW reconstructed Fermi surfaces. Dirac pockets have linear dispersion, as a consequence of which, it is possible to fulfil QLM condition because energy level splitting for linear band (Dirac States) is proportional to the square root of magnetic field ($\Delta_{LL}= \pm v_F(2e\hbar H)^{1/2}$), whereas, for parabolic bands, it is proportional to the magnetic field ($\Delta_{LL}= eH\hbar/m^*$ [16]).

Doubts have been raised on the applicability of QLM model to iron based superconductors [12, 64]. For instance, in Ba(Fe$_{1-x}$Co$_x$)$_2$As$_2$ the coefficient of linear magnetoresistance determined from experiments was not compatible with the QLM model [12]. QLM model also cannot explain LMR recently discovered in the high temperature paramagnetic phase of several electron doped compositions of the Ba(Fe$_{1-x}$Co$_x$)$_2$As$_2$ series [64], and the same can be said for the composition in present study. This is because it is highly unlikely that quantum limit conditions can be reached at such high temperatures and typical laboratory fields (~9T) in the absence of small Dirac pockets.

Another relevant model is due to Koshelev [73] which can in principle explain LMR in the SDW state of FeSCs. It is known that the SDW ordering leads to reconstruction of the Fermi surface. They argue that the scattering is strongest at the points on the Fermi surface which are connected by the nesting wave vector $Q_{AF}$. The area of regions close to these points grows linearly with the magnetic field, as a consequence of which, linear magnetoresistance is observed. This model also predicts a crossover between the linear and quadratic regimes of magnetoresistance at approximately 2 T, which is in agreement with the experiments in the SDW state. This model however has its own limitations in that, it has no mechanism which can explain the LMR in the paramagnetic regime of the optimally doped composition in present study, which has no SDW order.

## IV. Conclusions

In summary, we have carried out magnetotransport measurements and the scaling analysis in the optimally electron doped $Sr(Fe_{1-x}Co_x)_2As_2$ (x = 0.12) system. We observed that both the Kohler's and modified Kohler's scalings do not work for this system. Interestingly, Hall angle displays quadratic temperature dependence. These observations imply that the Hall angle and magnetoresistance are not governed by the same scattering process. We also observed linear magnetoresistance in this system. This suggests that linear magnetoresistance is possibly a generic feature of the paramagnetic phase of the electron doped 122 systems.

## V. Acknowledgements

The authors acknowledge Surjeet Singh for support during the course of this work. Rohit Kumar would like to acknowledge Rudra Prasad Jena and Sumesh Rana for technical support at UGC-DAE CSR, Indore and Jitender Kumar at IISER Pune for important suggestions regarding data analysis. R.K. and L. H. acknowledge support through DST-SERB Grant No. SR/FTP/PS-037/2010.

## VI. References

[1] Kamihara Y, Watanabe T, Hirano M and Hosono H 2008 Journal of the American Chemical Society **130** 3296–3297

[2] Paglione J and Greene R L 2010 Nature physics **6** 645

[3] Johnston D C 2010 Advances in Physics **59** 803–1061

[4] Stewart G 2011 Reviews of Modern Physics **83** 1589

[5] Nakajima Y, Taen T and Tamegai T 2009 Journal of the Physical Society of Japan **78** 023702

[6] Fang L, Luo H, Cheng P, Wang Z, Jia Y, Mu G, Shen B, Mazin I, Shan L, Ren C et al. 2009 Physical Review B **80** 140508

[7] Kasahara S, Shibauchi T, Hashimoto K, Ikada K, Tonegawa S, Okazaki R, Shishido H, Ikeda H, Takeya H, Hirata K et al. 2010 Physical Review B **81** 184519

[8] Pallecchi I, Bernardini F, Tropeano M, Palenzona A, Martinelli A, Ferdeghini C, Vignolo M, Massidda S and Putti M 2011 Physical Review B **84** 134524

[9] Kuo H H, Chu J H, Riggs S C, Yu L, McMahon P L, De Greve K, Yamamoto Y, Analytis J G and Fisher I R 2011 Physical Review B **84** 054540

[10] Chong S, Williams G, Kennedy J, Fang F, Tallon J and Kadowaki K 2013 EPL (Europhysics Letters) **104** 17002

[11] Bhoi D, Mandal P, Choudhury P, Pandya S and Ganesan V 2011 Applied Physics Letters **98** 172105

[12] Moseley D, Yates K, Peng N, Mandrus D, Sefat A S, Branford W and Cohen L 2015 Physical Review B **91** 054512

[13] Torikachvili M, BudKo S, Ni N, Canfield P and Hannahs S 2009 Physical Review B **80** 014521

[14] Ishida S, Liang T, Nakajima M, Kihou K, Lee C, Iyo A, Eisaki H, Kakeshita T, Kida T, Hagiwara M et al. 2011 Physical Review B **84** 184514

[15] Tanabe Y, Huynh K K, Heguri S, Mu G, Urata T, Xu J, Nouchi R, Mitoma N and Tanigaki K 2011 Physical Review B **84** 100508

[16] Huynh K K, Tanabe Y and Tanigaki K 2011 Physical review letters **106** 217004

[17] Gillett J, Das S D, Syers P, Ming A K, Espeso J I, Petrone C M and Sebastian S E 2010 arXiv preprint arXiv:1005.1330

[18] Hu R, Budko S L, Straszheim W E and Canfield P C 2011 Physical Review B **83** 094520

[19] Chu J H, Analytis J G, Kucharczyk C and Fisher I R 2009 Physical Review B **79** 014506

[20] Canfield P, Budko S, Ni N, Yan J and Kracher A 2009 Physical Review B **80** 060501

[21] Sebastian S E, Gillett J, Harrison N, Lau P, Singh D J, Mielke C and Lonzarich G 2008 Journal of Physics: Condensed Matter **20** 422203

[22] Sutherland M, Hills D, Tan B, Altarawneh M, Harrison N, Gillett J, OFarrell E, Benseman T, Kokanovic I, Syers P et al. 2011 Physical Review B **84** 180506

[23] Lee C H, Iyo A, Eisaki H, Kito H, Teresa Fernandez-Diaz M, Ito T, Kihou K, Matsuhata H, Braden M and Yamada K 2008 Journal of the Physical Society of Japan **77** 083704

[24] Kuroki K, Usui H, Onari S, Arita R and Aoki H 2009 Physical Review B **79** 224511

[25] Tegel M, Rotter M, Weiss V, Schappacher F M, Pöttgen R and Johrendt D 2008 Journal of Physics: Condensed Matter **20** 452201

[26] Krellner C, Caroca-Canales N, Jesche A, Rosner H, Ormeci A and Geibel C 2008 Physical Review B **78** 100504

[27] Nair S, Wirth S, Friedemann S, Steglich F, Si Q and Schofield A 2012 Advances in Physics **61** 583–664

[28] Nakajima Y, Shishido H, Nakai H, Shibauchi T, Behnia K, Izawa K, Hedo M, Uwatoko Y, Matsumoto T, Settai R et al. 2007 Journal of the Physical Society of Japan **76** 024703

[29] Nakajima Y, Shishido H, Nakai H, Shibauchi T, Hedo M, Uwatoko Y, Matsumoto T, Settai R, Onuki Y, Kontani H et al. 2008 Physical Review B **77** 214504

[30] Kohler M 1938 *Annalen der Physik* **424** 211–218

[31] Hurd C 2012 *The Hall effect in metals and alloys* (Springer Science & Business Media)

[32] Pippard A B 1989 *Magnetoresistance in metals* Vol. 2 (Cambridge University Press)

[33] Kimura T, Miyasaka S, Takagi H, Tamasaku K, Eisaki H, Uchida S, Kitazawa K, Hiroi M, Sera M and Kobayashi N 1996 Physical Review B **53** 8733

[34] Harris J, Yan Y, Matl P, Ong N, Anderson P, Kimura T and Kitazawa K 1995 Physical review letters **75** 1391

[35] Wang L, Wang C Y, Sou U C, Yang H C, Chang L, Redding C, Song Y, Dai P and Zhang C 2013 Journal of Physics: Condensed Matter **25** 395702

[36] Eom M, Na S, Hoch C, Kremer R and Kim J 2012 Physical Review B **85** 024536

[37] Cheng P, Yang H, Jia Y, Fang L, Zhu X, Mu G and Wen H H 2008 Physical Review B **78** 134508

[38] Gnida D, Matusiak M and Kaczorowski D 2012 Physical Review B **85** 060508

[39] Abe Y, Ando Y, Takeya J, Tanabe H, Watauchi T, Tanaka I and Kojima H 1999 Physical Review B **59** 14753

[40] Konstantinovic Z, Laborde O, Monceau P, Li Z and Raffy H 1999 Physica B: Condensed Matter **259** 569–570

[41] Balakirev F, Trofimov I, Guha S, CieplakMZ and Lindenfeld P 1998 Physical Review B **57** R8083

[42] Morozova N V, Karkin A E, Ovsyannikov S V, Umerova Y A, Shchennikov V V, Mittal R and Thamizhavel A 2015 Superconductor Science and Technology **28** 125010

[43] Ping Z, Gen-Fu C, Zheng L,Wan-Zheng H, Jing D, Gang L, Nan-Lin W and Jian-Lin L 2009 Chinese Physics Letters **26** 107401

[44] Harnagea L, Mani G, Kumar R and Singh S 2018 Physical Review B **97** 054514

[45] Kim J S, Khim S, Yan L, Manivannan N, Liu Y, Kim I, Stewart G and Kim K H 2009 Journal of Physics: Condensed Matter **21** 102203

[46] Aswartham S, Nacke C, Friemel G, Leps N, Wurmehl S, Wizent N, Hess C, Klingeler R, Behr G, Singh S et al. 2011 Journal of Crystal Growth **314** 341–348

[47] Ni N, Tillman M, Yan J Q, Kracher A, Hannahs S, BudKo S and Canfield P 2008 Physical Review B **78** 214515

[48] Kasinathan D, Ormeci A, Koch K, Burkhardt U, SchnelleW, Leithe- Jasper A and Rosner H 2009 New journal of physics **11** 025023

[49] Allred J M, Taddei K M, Bugaris D E, Avci S, Chung D Y, Claus H, dela Cruz C, Kanatzidis M, Rosenkranz S, Osborn R et al. 2014 Physical Review B **90** 104513

[50] Avci S, Chmaissem O, Chung D Y, Rosenkranz S, Goremychkin E A, Castellan J P, Todorov I S, Schlueter J A, Claus H, Daoud-Aladine A et al. 2012 Physical Review B **85** 184507

[51] Rullier-Albenque F, Colson D, Forget A and Alloul H 2009 Physical review letters **103** 057001

[52] Liu Y and Lograsso T A 2014 Physical Review B **90** 224508

[53] Kontani H 2008 Reports on Progress in Physics **71** 026501

[54] Luo N and Miley G 2002 Physica C: Superconductivity **371** 259–269

[55] Brouet V, Lin P H, Texier Y, Bobroff J, Taleb-Ibrahimi A, Le F`evre P, Bertran F, Casula M, Werner P, Biermann S et al. 2013 Physical review letters **110** 167002

[56] Koshelev A 2016 Physical Review B **94** 125154

[57] Breitkreiz M, Brydon P and Timm C 2014 Physical Review B **89** 245106

[58] Breitkreiz M, Brydon P and Timm C 2013 Physical Review B **88** 085103

[59] Anderson P 1991 Physical review letters **67** 2092

[60] Hlubina R and Rice T 1995 Physical Review B **51** 9253

[61] Stojkovic-acute B P and Pines D 1997 Physical Review B **55** 8576

[62] Kokanović I, Cooper J, Naqib S, Islam R and Chakalov R 2006 Physical Review B **73** 184509

[63] Ando Y and Murayama T 1999 Physical Review B **60** R6991

[64] Kumar R, Singh S and Nair S 2018 arXiv preprint arXiv:1801.03768

[65] Kontani H, Kanki K and Ueda K 1999 Physical Review B **59** 14723

[66] Kontani H 2001 Journal of the Physical Society of Japan **70** 1873–1876

[67] Kontani H 2002 Physica B: Condensed Matter **312** 25–27

[68] Richard P, Nakayama K, Sato T, Neupane M, Xu Y M, Bowen J, Chen G, Luo J, Wang N, Dai X et al. 2010 Physical review letters **104** 137001

[69] Terashima T, Hirose H T, Graf D, Ma Y, Mu G, Hu T, Suzuki K, Uji S and Ikeda H 2018 Physical Review X **8** 011014

[70] Abrikosov A 1998 Physical Review B **58** 2788

[71] Abrikosov A 2000 EPL (Europhysics Letters) **49** 789

[72] Abrikosov A 2003 Journal of Physics A: Mathematical and General **36** 9119

[73] Koshelev A 2013 Physical Review B **88** 060412

## VII. Figures

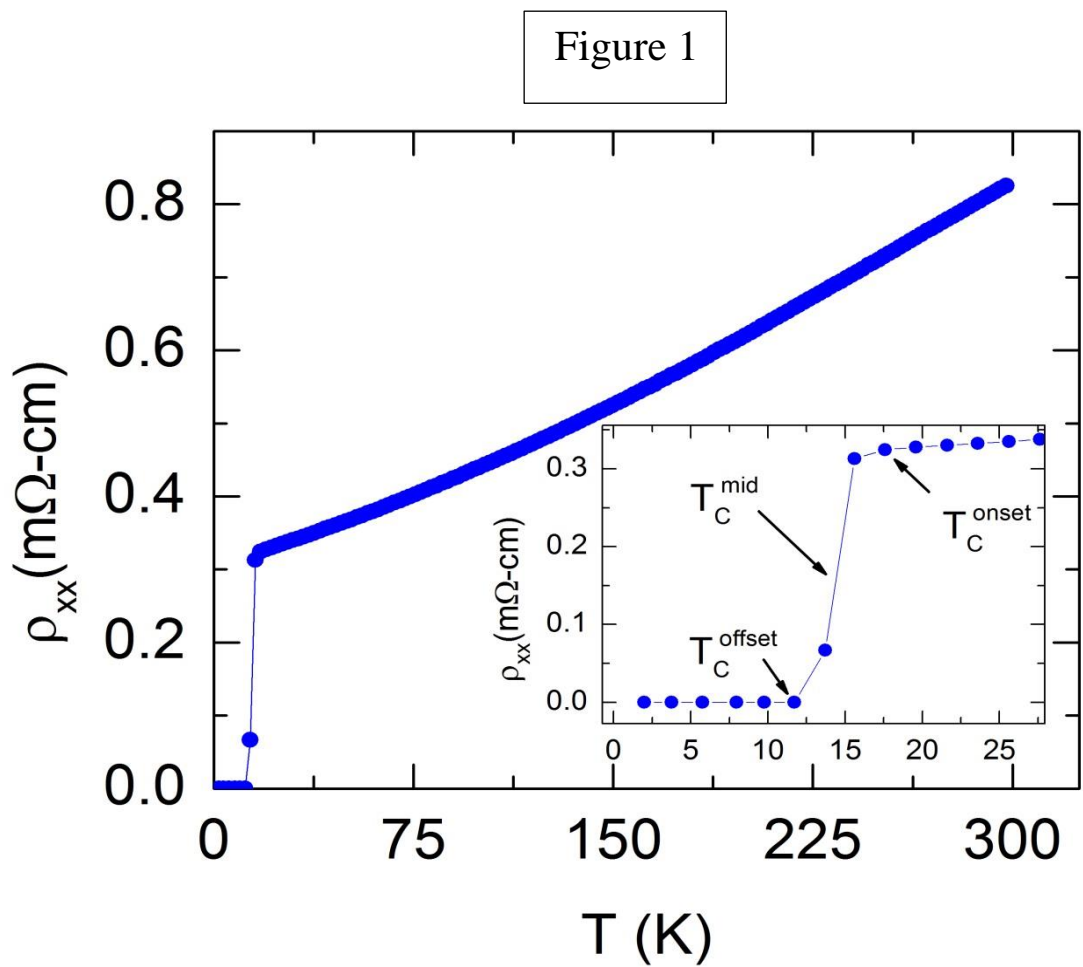

Figure 1

Figure 1. Zero field in plane resistivity from 2-300 K. Inset shows the resistivity near superconducting transition. $T_C$ is assigned at the midpoint of the transition which is ≈14.5 K. $T_C^{onset}$ marks the beginning of the superconducting transition as indicated by an arrow. $T_C^{offset}$ is the temperature at which resistivity reduces to zero. $T_C$ is approximately the average of $T_C^{onset}$ and $T_C^{offset}$.

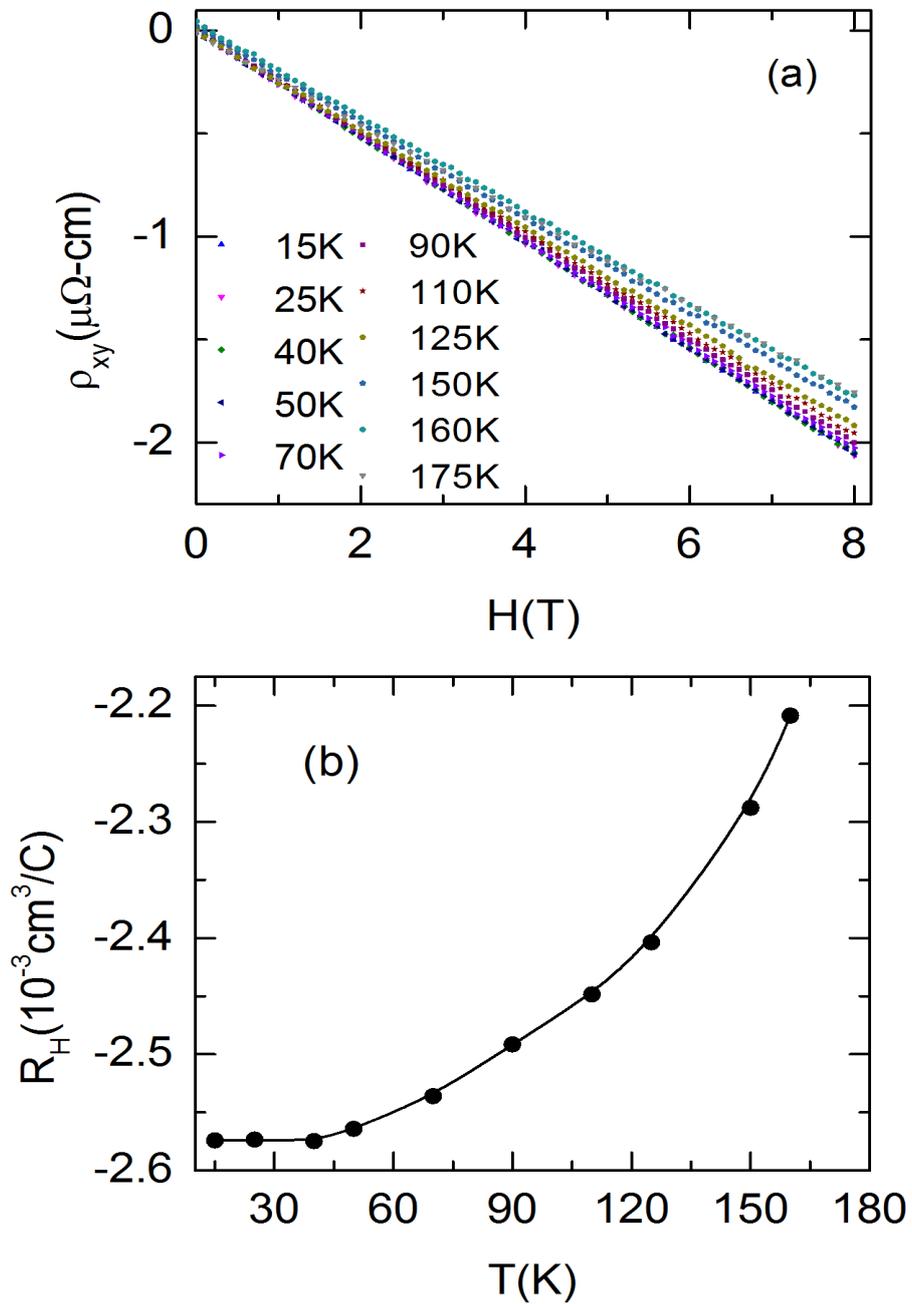

Figure 2. Hall resistivity ($\rho_{xy}$) as a function of magnetic field (a). In (b) Hall coefficient derived by dividing Hall resistivity from magnetic field ($R_H = \rho_{xy}/H$) is shown from 15-160 K

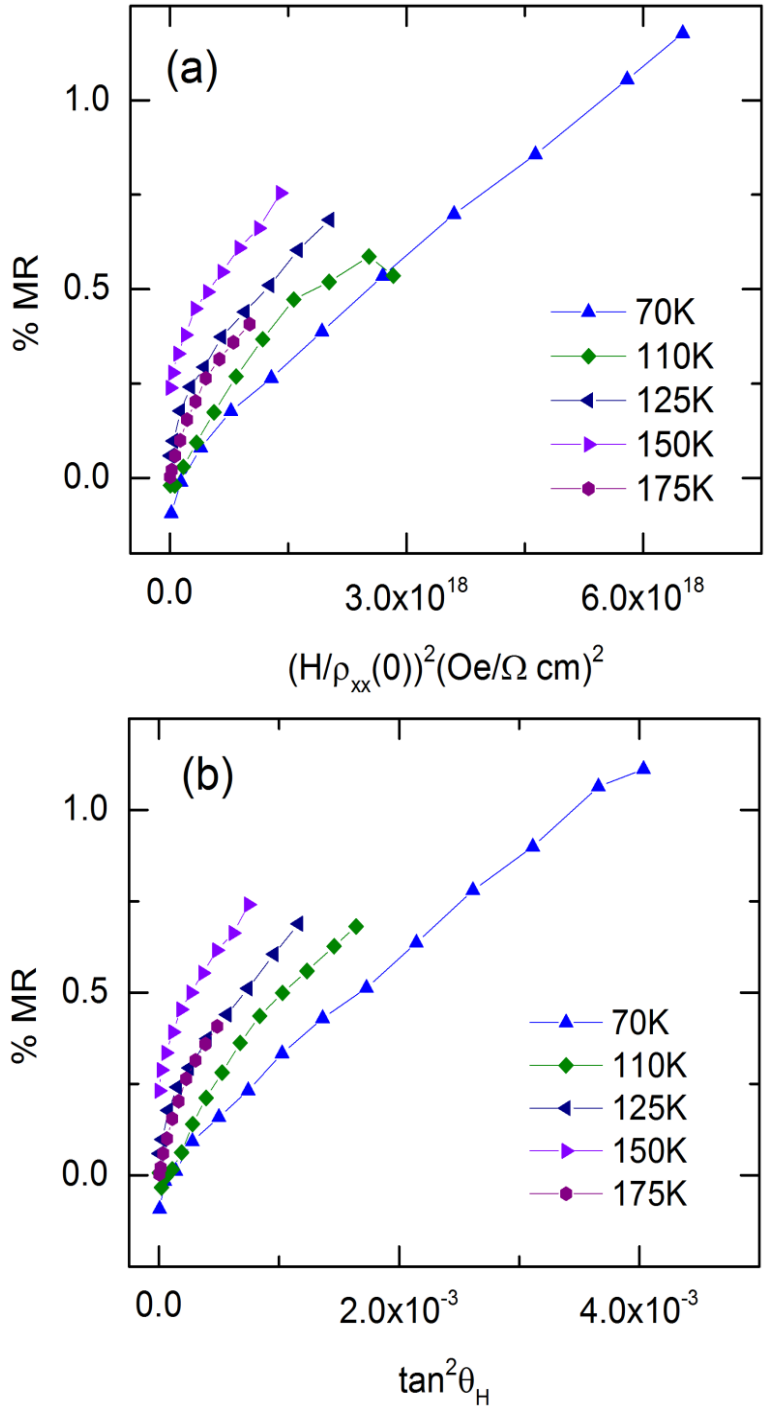

Figure 3. (a) and (b) shows scaling of magnetoresistance in terms of Kohler and modified Kohler's rule respectively. The magnetic field is in the range from 0-8 T in both (a) and (b).

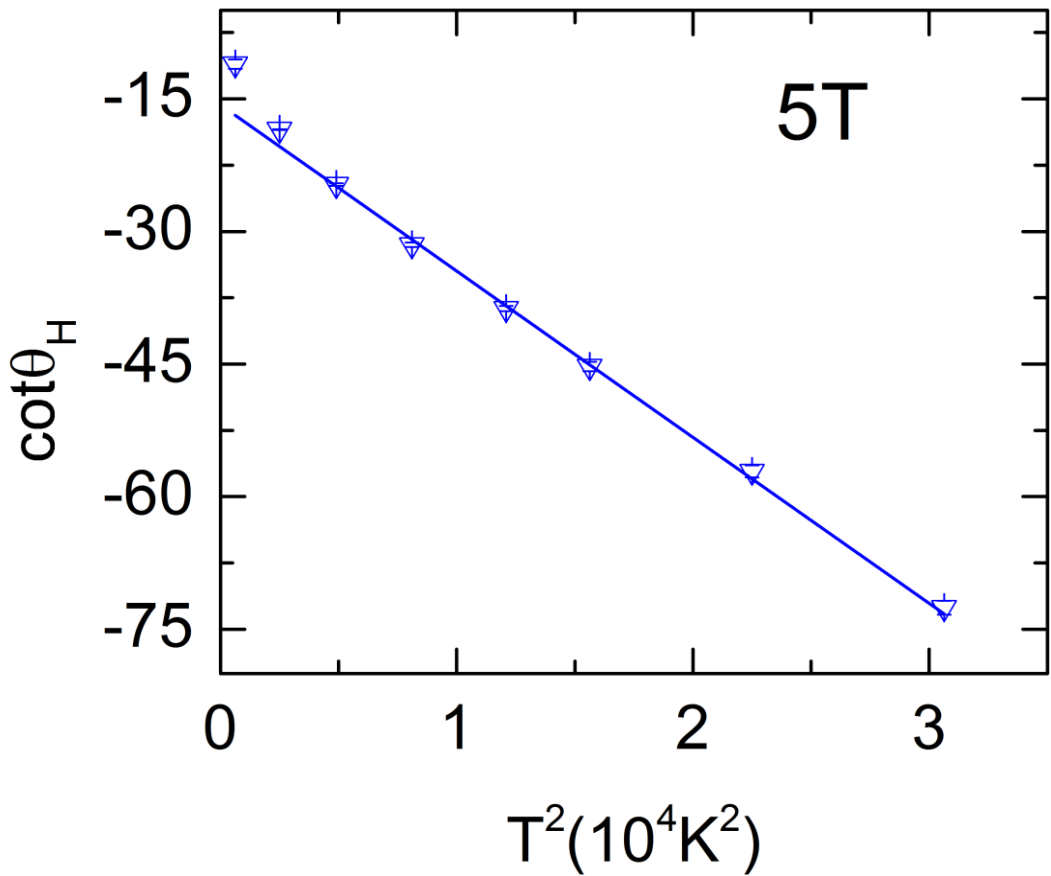

Figure 4. $T^2$ dependence of Hall angle ($cot\theta_H \propto T^2$) calculated at 5 Tesla is shown from 25-175 K.

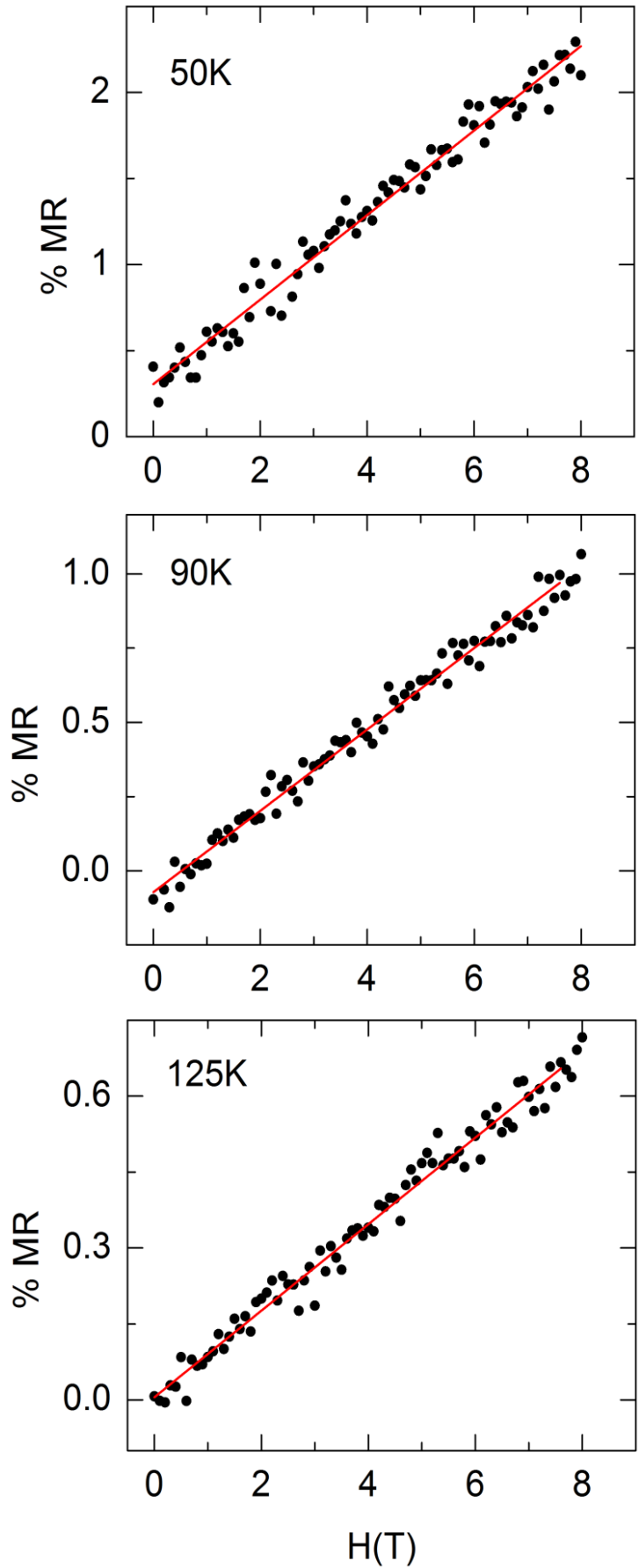

Figure 5. Linear magnetoresistance at three representative temperatures. Solid lines are linear fit to the data.